\documentclass[aps,prl,twocolumn,floats,showpacs]{revtex4}
\usepackage{epsfig}
\usepackage{graphicx}
\usepackage{amsmath}

\begin{document}
 
\title{Two-Step Enantio-selective Optical Switch}

\author
{Petr Kr\'al$^1$, Ioannis Thanopulos$^1$, Moshe Shapiro$^1$
and Doron Cohen$^2$}

\affiliation{$^1$Department of Chemical Physics, Weizmann Institute of Science,
         Rehovot, Israel}

\affiliation{$^2$Department of Physics, Ben-Gurion University, 
Beer-Sheva, Israel}

\date{\today}

\begin{abstract}
We present an optical ``enantio-selective switch", that, in two steps, 
turns a (``racemic") mixture of left-handed and right-handed chiral 
molecules into the enantiomerically pure state of interest.
The optical switch is composed of an ``enantio-discriminator" and an
``enantio-converter" acting in tandem. The method is robust, insensitive 
to decay processes, and does not require molecular preorientation. We
demonstrate the method on the purification of a racemate of (transiently 
chiral) D$_2$S$_2$ molecules, performed on the nanosecond timescale.
\end{abstract}

 
\maketitle


Asymmetric synthesis \cite{Nobel} and chiral purification
\cite{Piezo-dis,Chem-dis,Lahav} of a (``racemic") mixture
of enantiomers (chiral molecules and their mirror images) are
among the most important and difficult tasks in chemistry. 
The possibility of achieving purification solely by optical
means has been also theorized \cite{bs91,salam98,fujimanz,MosheEB}.
In particular, this goal could be realized via a ``laser distillation" 
scheme \cite{MosheEB}, in which a repetitive use of three light pulses 
\cite{Dipoles} gradually purifies the system.

Recently, we suggested a method for achieving chiral separation, termed 
``Cyclic Population Transfer" (CPT) \cite{CPT}. The approach is akin to the
Adiabatic Passage (AP) \cite{Grisch,oreg,Bergmann}, used to {\it completely} 
transfer population between quantum states, that are usually optically coupled
as, $|1 \rangle \leftrightarrow |2 \rangle \leftrightarrow |3 \rangle$. In
chiral molecules, lacking an inversion center and thus having eigenstates
with ill defined parity, it is possible to close the ``cycle" by introducing 
a third field which couples the states $|1 \rangle\leftrightarrow |3 \rangle$ 
directly. The interference of one and two-photon transitions along the
two paths renders the evolution, in this CPT scheme, dependent on the 
total {\it phase} $\varphi$ of the three (material+optical) coupling terms
\cite{CPT}. Since the transition dipoles of the two enantiomers differ in 
sign, the evolution in the two under the action of the three fields is 
different, and the enantiomers can be {\it separated}. In particular,
if they initially occupy state $|1 \rangle$, one enantiomer can be excited 
to state $|2 \rangle$, while the other is transferred to state 
$|3 \rangle$, or vice versa, depending on the phase $\varphi$.

The great advantage of the CPT scheme is that the separation process
can be completed in just one step. We can do it using optical transitions, 
taking place on the ground electronic surface, thereby avoiding
disruptive competing processes, such as dissociation and internal
conversion \cite{MosheEB}. However, CPT is not particularly robust 
with respect to variation of the laser pulses and it does {\it not} 
convert one enantiomer into another.

In this work, we demonstrate that optical chiral purification can be 
accomplished in just two (separation and conversion) steps, in a 
scheme which overcomes the above drawbacks. In the first one, called
``enantio-discriminator", we excite one enantiomer while leaving the other
in its initial state.  In the second one, called ``enantio-converter",
the enantiomer excited in the previous step is converted to its mirror-image 
form. Thus, in just two steps, the racemic mixture of chiral molecules 
could be converted into the enantiomer of choice.

\begin{figure}
\vspace*{-12mm}
\begin{center}
\leavevmode
 \hbox{\epsfxsize=80mm \epsffile{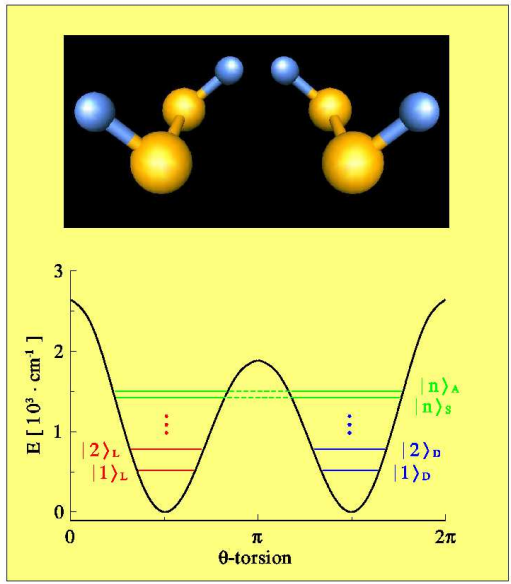} }
\vspace*{-15mm}
\end{center}
\caption{The left and right D$_2$S$_2$ enantiomers, which undergo 
stereomutation about the $S-S$ bond. Below is shown the 
double well potential energy for their torsional motion on the ground 
electronic state. In the considered $100$ ns timescale, the lower energy 
(red and blue) ro-vibronic states, schematically shown here, have a 
definite (left and right) 
symmetry, while at higher energies they become split into symmetric and 
antisymmetric (green) states.}
\label{CUBEPOT}
\end{figure}

We present the enantiomer switch on the (transiently chiral) D$_2$S$_2$ 
molecule, shown schematically in Fig.~\ref{CUBEPOT}.
The molecule has 6 vibrational degrees of freedom, with the large-amplitude
torsional motion of the D (blue) atoms about the ${\rm S-S}$ (orange) bond.
A one-dimensional cut of the ground electronic potential energy surface 
along the enantio-mutative path is shown in Fig.~\ref{CUBEPOT}.  From our 
{\it ab-initio} calculation, we have determined that the enantio-mutation 
of the molecules can be realized through a {\it cis} barrier ($\theta=0$), 
2700~cm$^{-1}$ in height, or the {\it trans} barrier ($\theta=\pi$), 
1900~cm$^{-1}$ in height.
The calculated tunneling splitting of the lowest torsional states
gives enantiomeric lifetimes of several msec, in accordance with 
previous reports on mode selective stereomutation \cite{Quack}. Thus, 
although ${\rm D_2S_2}$ is not a chiral molecule in the conventional 
sense, molecular configurations described by superpositions of the lowest
torsional states, localized in one minimum of the double well potential,
stay chiral for sufficiently long times to be detected.

The two-step process makes use of five pairs of D$_2$S$_2$ ro-vibrational 
eigenstates. The vibrational states correspond to the combined torsional and
${\rm S-D}$ asymmetric stretching modes. The rotational states 
correspond to a rigid rotor.  Within the pairs used \cite{STATES}, 
each eigenstate has an $S/A$ (symmetric/antisymmetric) label denoting 
its symmetry with respect 
to inversion, and the chiral states with $L/D$ labels are composed as,
$|k\rangle_{L,D}=\frac{1}{\sqrt{2}}\, \bigl(\, |k\rangle_S \pm
|k\rangle_A \, \bigr)\, , \ k=1,\ldots,4$. The $|k\rangle_L
\leftrightarrow|k\rangle_D$ interconversion period is 
$\tau_s \approx 33,~3.3,~0.165$ ms for $k=1,2,3$, respectively, and 
$\tau_s \approx 0.05~\mu$s for $k=4$. The higher lying $|5 \rangle_S$ 
and $|5 \rangle_A$ states are separated by $\Delta E_{S,A}^5=0.38$~cm$^{-1}$, 
for which $\tau_s\approx 0.1$~ns, and can be thus separately addressed by 
$ns$ pulses, that are strong enough, to ensure adiabaticity.  The ns pulses 
are also long enough to address individual lower lying ro-vibronic levels,
but, since $\tau\ll \tau_s$, they cannot separately address the narrowly 
split $S$ and $A$ states, so the symmetry-broken combination states 
$|k\rangle_{L,D}$ ($k=1-4$) become physically meaningful.

We now describe in details the dynamics of the switch. Denoting the energy 
of level $|i \rangle$ by $\omega_i$ ($\hbar=1$ in atomic units), we 
choose the external electric field 
to be a sum of components, each being in {\it resonance} with one of the
$|i\rangle\leftrightarrow|j\rangle$ transition frequencies of interest,
${\bf E}(t)  =  \sum_{i\ne j} {\cal R}_e\Big[\hat\epsilon\,
{\cal E}_{i,j}(t)e^{-i\omega_{i,j}t}\Big]$, where $\omega_{i,j}= \omega_i - 
\omega_j$, and $\hat\epsilon$ is the polarization direction. The Hamiltonian 
of the system in the rotating wave approximation is,
\begin{eqnarray}
{\sf H}= \sum_{i=1}^N\omega_i |i\rangle \langle i|
    + \sum_{i>j=1}^N \bigl( \Omega_{i,j}(t)
      e^{-i\omega_{i,j}t} |i\rangle \langle j| + {\rm H.c.}\bigr)\, .
\label{H}
\end{eqnarray}
It depends on the Rabi frequencies, $\Omega_{i,j}(t)=\mu_{i,j}\, {\cal E}_{i,j}
(t)$, where $\mu_{i,j}$ are the transition-dipole matrix elements. 
Expanding the system wave function in the material states $|i \rangle$ as,
$|\psi(t)\rangle = \sum_{i=1}^N c_i(t)\, e^{-i\omega_i t}~|i \rangle$, 
the (column) vector of the slow varying coefficients ${\bf c}=
(c_1, c_2,..., c_N)^{\sf T}$, with ${\sf T}$ designating the matrix
transpose, is the solution of the matrix-Schr\"{o}dinger equation
$\dot{\bf c}(t)=-i\, {\sf H}(t) \cdot{\bf c}(t)$, where ${\sf H}(t)$
is an effective Hamiltonian matrix, given explicitly 
below for the two processes.

The scheme of the three-level ``enantio-discriminator" is shown in the 
upper panel of Fig.~2. 
Assuming that the system is at low temperature, so that we can practically
start with a mixture of chiral ${\rm D_2S_2}$ molecules in the ground 
$|1\rangle_L$ and $|1\rangle_D$ states, the task of the discriminator 
is to selectively transfer one enantiomer to the $|3\rangle$ state and 
to keep the other in the
$|1\rangle$ state. Due to the degeneracy of the $|i\rangle_L$
and the $|i\rangle_D$ levels ($i=1,2,3$), the field
${\bf E}(t)$ simultaneously excites the resonant $|i\rangle_{L,D} 
\leftrightarrow |j\rangle_{L,D}~,i\ne j=1,2,3$ transitions of 
both enantiomers \cite{Dipoles}.

In the first enantio-discriminator step, the effective Hamiltonian matrix is,
\begin{equation}
{\sf H}(t)= \left[
\begin{array}{ccc}
0 & \Omega_{1,2}^\ast(t) & \Omega_{1,3}^\ast(t) \\
  \Omega_{1,2}(t) & 0 & \Omega_{2,3}^\ast(t) \\
  \Omega_{1,3}(t) & \Omega_{2,3}(t) & 0
\end{array}
\right]\ .
\label{HamSE}
\end{equation}
The phases of the Rabi frequencies $\Omega_{i,j}(t)$
are given as in the CPT scheme by $\phi_{i,j}=\phi^\mu_{i,j}+\phi^E_{i,j}$,
where $\phi^\mu_{i,j}$ are the phases of the dipole matrix elements
$\mu_{i,j}$, and $\phi^E_{i,j}$ are the phases of the electric field
components ${\cal E}_{i,j}$. The evolution of the system is determined 
\cite{CPT} by the {\it total} phase $\varphi\equiv \phi_{1,2}+\phi_{2,3} 
+\phi_{3,1}$.  This is most noticeable at the time $t=\tau$, for which 
the three Rabi frequencies are equal in magnitude, $|\Omega_{1,2}|
=|\Omega_{1,3}|=|\Omega_{2,3}|=\Omega$. 
Denoting the eigenvalues of the Hamiltonian of Eq.~(\ref{HamSE})
as $E_{-}$, $E_{0}$ and $E_{+}$, it is easy to show that 
they exhibit exact degeneracies (crossings) at $t=\tau$, with
$E_{+}=2\, \Omega$ and $E_{-}=E_{0}=2\, \Omega\, \cos(2\pi/3)$, 
for $\varphi=0$, and $E_{-}=-2\, \Omega$ and $E_{+}=E_{0}=-2\, \Omega\, 
\cos(2\pi/3)$, for $\varphi=\pi$. 
Depending on the polarizations of the fields, one or all three Rabi
frequencies $\Omega_{i,j}$ of the two enantiomers differ by a sign
\cite{CPT}, making $\varphi$ {\it differ} by $\pi$ for the two. 
Therefore, the above two degeneracies (crossings) occur at 
different enantiomers, leading subsequently to their totally
different dynamics, depicted in the lower panel of Fig.~2.

The overall enantio-discriminator works as follows: We start with a 
``dump" pulse ${\cal E}_{2,3}(t)$ that couples the $|2\rangle$ and 
$|3\rangle$ states and has the Rabi frequency 
$\Omega_{2,3}(t)=\Omega^{\rm max}\, f(t)$, where $\Omega^{\rm max} =1$ 
ns$^{-1}$ and $f(t)= \exp[-t^2/\tau^2]$. At this stage of the process 
all the population resides in the $|E_{0}\rangle$ (adiabatic) eigenstate.
In the second stage we {\it simultaneously} add two ``pump" pulses
of the Rabi frequencies $\Omega_{1,2}(t) =\Omega_{1,3}(t)=\Omega^{\rm max}\, 
f(t-2\,\tau)$, that couple the $|1\rangle \leftrightarrow |2\rangle$ and the 
$|1\rangle \leftrightarrow |3\rangle$ states.  We choose the phases of the 
optical fields such that $\varphi=0$ for one enantiomer and, inevitably,
$\varphi=\pi$ for the other. Therefore, the population, which has been 
following in both enantiomers the initial adiabatic level $|E_{0}\rangle$, 
goes at $t=\tau$ smoothly through the crossing region and {\it diabatically} 
transfers to either the $|E_{-}\rangle$ or the $|E_{+}\rangle$ states, 
depending on whether $\varphi=0$ or $\varphi=\pi$, i.e., on the identity 
of the enantiomer.

After the crossing is complete, at $t>\tau$, the process becomes
adiabatic again, with the enantiomer population residing fully in
either $|E_{-}\rangle$ or $|E_{+}\rangle$. At this stage we slowly 
switch off the ${\cal E}_{1,2}(t)$ pulse while making sure that
the ${\cal E}_{1,3}(t)$ field remains on. This is done by choosing
$\Omega_{1,3}(t) = \Omega^{\rm max}\bigl(f(t-2\,\tau)+
f(t-4\,\tau) \exp\{-i\, t\, \Omega^{\rm max}\, f(t-6\,\tau)\} \bigr)$.
As a result, the zero adiabatic eigenstate $|E_0\rangle$ correlates
adiabatically with state $|2\rangle$, which thus becomes {\it empty}
after this process, while the occupied $|E_+\rangle$ and $|E_-\rangle$
states correlate to, $|E_{\pm}\rangle\rightarrow
\left(|1\rangle \pm |3\rangle\right)/\sqrt{2}$.

\begin{figure}
\begin{center}
\leavevmode
\vspace*{-2mm}
\hspace*{-37.5mm}
{ \hbox{\epsfxsize=160mm \epsffile{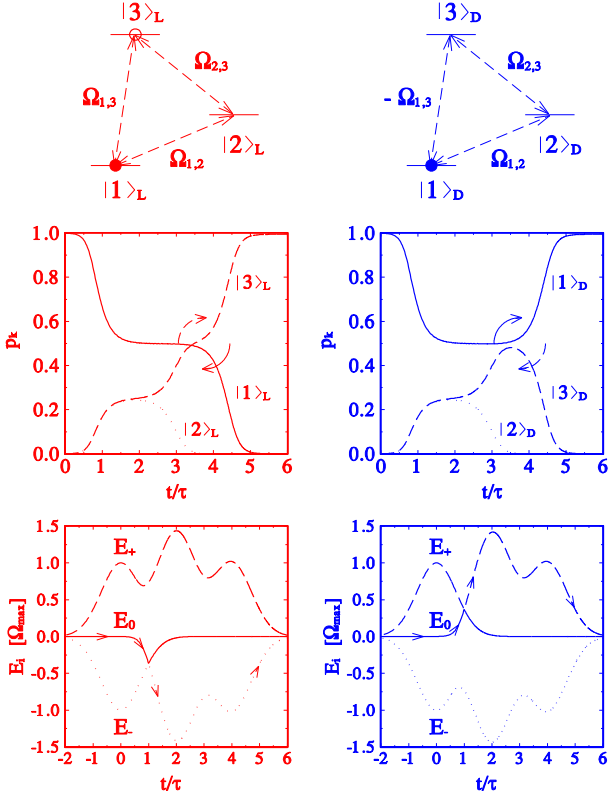}}}
\vspace*{-20mm} 
\end{center}
\caption{(Upper panel) A schematic plot of the enantio-discriminator.
The three levels of each enantiomer are resonantly coupled by three
fields. (Middle panel) The time evolution of the population of the three 
levels. Both enantiomers start in the $|1\rangle$ state. At the end 
of the process the $L$ enantiomer is transferred to the $|3\rangle_L$ 
state, while the $D$ enantiomer remains in the initial $|1\rangle_D$ 
state. (Lower panel) The time-dependence of the 
eigenvalues of the Hamiltonian of Eq. (\ref{HamSE}).
The population initially follows the $|E_0 \rangle$ dark state.
At $t\approx \tau$ the population crosses over diabatically to
$|E_- \rangle$ for one enantiomer and to $|E_+ \rangle$ for the other.}
\label{CUBERL}
\end{figure}

The {\it chirp}, $\exp\{-i\, t\, \Omega^{\rm max}\, f(t-6\,\tau)\}$,
in the second term of $\Omega_{1,3}(t)$ causes a $\pi/2$ {\it rotation} 
in the $\{ |1 \rangle, |3 \rangle \}$ subspace at $t\approx 5\,\tau$. As 
a result, state $|E_{+}\rangle$ goes over to state $|3\rangle$ and 
state $|E_{-}\rangle$ goes over to state $|1\rangle$, or vice versa,
depending on $\varphi$.  The net result of the adiabatic passage and 
the rotation is that one enantiomer returns to its initial 
$|1\rangle$ state and the other switches over to the $|3\rangle$ state.
As shown in the middle panel of Fig.~2, the enantio-discriminator
is very {\it robust}, in contrast to the CPT scheme \cite{CPT}, with all 
the population transfer processes occurring in a smooth fashion.

Assuming that the $L$ enantiomer has been excited to the $|3\rangle_L$ state,
we now proceed to convert it into a $D$ enantiomer in the $|4\rangle_D$ state,
by going through a linear superposition of $|5\rangle_{S}$ and $|5\rangle_{A}$ 
states, while leaving intact the $D$ enantiomer in the state $|1\rangle_D$.
This ``enantio-converter" process, based on a new multi-path transfer 
technique \cite{SAPON},  thus schematically follows the pathway $|3\rangle_L 
\rightarrow \alpha\, e^{-i\, \omega_{5S} t}\, |5\rangle_{S} + \beta\, 
e^{-i\, \omega_{5A} t}\, |5\rangle_{A} \rightarrow |4\rangle_D$, shown in 
Fig.~3. The transfer is realized by simultaneously introducing two ``dump" 
pulses ${\cal E}_{4,5S}(t)$ and ${\cal E}_{4,5A}(t)$ (of duration $\tau \gg
(\omega_{5S}-\omega_{5A})^{-1}$), which resonantly
couple each of the $|5\rangle_{S}$ and $|5\rangle_{A}$ states to the 
$|4\rangle_L$ and $|4\rangle_D$ state.  After a delay of $2\, \tau$, we 
introduce two (``pump") pulses ${\cal E}_{3,5S}(t)$ 
and ${\cal E}_{3,5A}(t)$, which resonantly couple each of the 
$|5\rangle_{S}$ and $|5\rangle_{A}$ states to the $|3\rangle_L$ and
$|3\rangle_D$ state.  In this process only the $|3\rangle_D$ 
and $|4\rangle_L$ states are populated, while the $|3 \rangle_L$ and 
$|4 \rangle_D$ states, degenerate with them, respectively, stay empty. 
This is because the empty pair of states is coupled to the $|5\rangle_{S}$ 
and $|5\rangle_{A}$ states by vectors of Rabi frequencies $\Omega_i$, that 
are {\it orthogonal} \cite{SAPON} to analogous vectors of the populated 
pair of states, respectively. The {\it symmetry conversion} of the excited
enantiomer is achieved by choosing $\Omega_{4,5S}(t)$ to have the same 
sign as $\Omega_{3,5S}(t)$ and $\Omega_{4,5A}(t)$ to have an opposite 
sign to $\Omega_{3,5A}(t)$. 

The enantio-converter is described by the Hamiltonian, 
\begin{equation}
{\sf H}= \left[
\begin{array}{cccccc}
0 & 0 &-\Omega_{3,5S} & \Omega_{3,5A} & 0 & 0 \\
0 & 0 & \Omega_{3,5S} & \Omega_{3,5A} & 0 & 0 \\
 -\Omega_{3,5S}^\ast & \Omega_{3,5S}^\ast & 0 & 0
&-\Omega_{4,5S}^\ast & \Omega_{4,5S}^\ast \\
  \Omega_{3,5A}^\ast & \Omega_{3,5A}^\ast & 0 & 0
& \Omega_{4,5A}^\ast & \Omega_{4,5A}^\ast \\
0 & 0 &-\Omega_{4,5S} & \Omega_{4,5A} & 0 & 0 \\
0 & 0 & \Omega_{4,5S} & \Omega_{4,5A} & 0 & 0 \\
\end{array}
\right] ~,\
\label{HamCA}
\end{equation}
with the time-dependent wavefunction given by the vector ${\bf c}(t)=
(c_{3L},c_{3R},c_{5S},c_{5A},c_{4L},c_{4D})$ of expansion coefficients 
in the $|i \rangle$ states.  The Hamiltonian matrix (\ref{HamCA}) has four 
non-zero eigenvalues and two null eigenvalues, $\lambda_{1,2}=0$, that 
correspond to two dark states with the coefficients
${\bf c}_{1}(t)= \bigl(-d_+, -d_-, 0,0, 2\, \Omega_{3,5S},0 \bigr)$,
${\bf c}_{2}(t)= \bigl(-d_-, -d_+, 0,0,0, 2\, \Omega_{3,5S} \bigr)$,
where $d_\pm\equiv\Omega_{4,5S} \pm r\,\Omega_{4,5A}$, with 
$r\equiv\Omega_{3,5S}/\Omega_{3,5A}$. These expressions show that the 
system can follow two possible paths, where only {\it one} of them is
flipping the symmetry of the initial state. Assuming, for simplicity, 
that $r=1$ and $r^{'} =\Omega_{4,5S} /\Omega_{4,5A}=1$, we find out that 
at beginning of the process only the dark state ${\bf c}_{1}(t_{ini})$
correlates with the initial state $|3\rangle_{L}$, {\it i.e.} the vector 
${\bf c}(t_{ini}) =(1,0,0,0,0,0)$. At the end of the processes, 
this dark state correlates with the vector ${\bf c}_{1}(t_{end})
=(0,0,0,0,1,0)$ for the $|4\rangle_{L}$ state, so the symmetry is 
preserved. On the other hand, if we flip the phase of just one dump or
one pump field component ($r=-1$ or $r^{'}=-1$) the system follows the
dark state ${\bf c}_{2}$, which correlates at the end with the state
${\bf c}_{2}(t_{end})=(0,0,0,0,0,1)$. The final population thus occupies
the $|4\rangle_{D}$ state, with the opposite symmetry. This brings the
whole population to a {\it single enantiomer form}.

\begin{figure}
\begin{center}
\leavevmode
\hspace*{-25mm}
{\vspace*{-2mm}
\hbox{\epsfxsize=130mm \epsffile{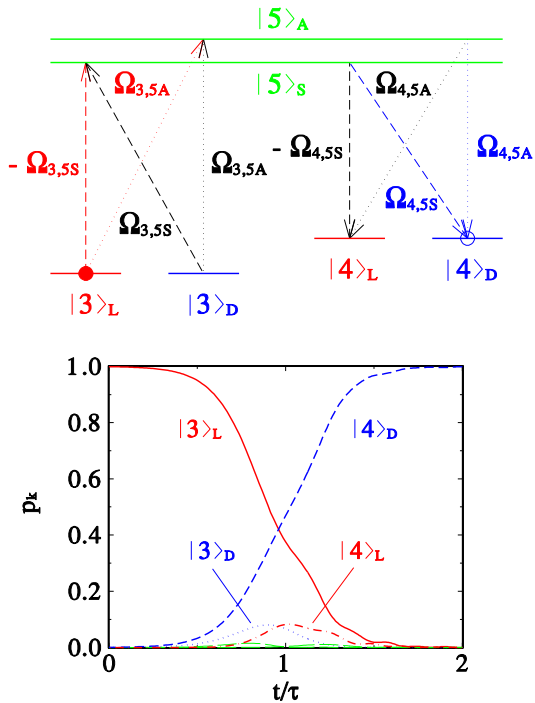} }
\vspace*{-10mm} }
\end{center}
\caption{(Upper plot) Scheme of the enantio-converter.  The population
passes from the $|3\rangle_{L}$ state to the $|4\rangle_{D}$ state, 
while going through the superposition of $|5\rangle_{S,A}$ states. 
(Lower plot) The time-dependent populations $p_i$ on the levels.}
\label{CUBECOSA}
\end{figure}

In Fig.~3, we show evolution of the calculated populations $p_i=|c_i|^2$.
The process starts in the $|3\rangle_{L}$ state and ends in the 
$|4\rangle_{D}$ state. The Rabi frequencies are
$\Omega_{3,5S}(t) = \Omega^{\rm max}\, f(t-2\, \tau)$,
$\Omega_{3,5A}(t) =0.5\  \Omega^{\rm max}\, f(t-2\, \tau)$,
$\Omega_{4,5S}(t) = 0.4\ \Omega^{\rm max}\, f(t)$,
$\Omega_{4,5A}(t) =-\Omega^{\rm max}\, f(t)$, with $\Omega^{\rm max}=30$ 
ns$^{-1}$ and $f(t)$ as in the discriminator. 
Notice, that we have rather different $r=2$,  $r^{'}=-0.4$, which
shows that this system is {\it robust}, {\it i.e.} it does not require that 
$|r|=1$ and $|r^{'}|=1$ to follow this complete-transfer path,
but these ratios need to have the right sign.

Because the higher excited $|5\rangle_{S,A}$ states never get populated in 
the conversion, the switch is immune to $T_2$-like dephasing processes, which 
destroy the relative phase between the components of excited superposition 
states.  For the same reason, it is insensitive to dissociation and/or internal 
conversion, even if {\it higher} electronic surfaces, having well separated 
$|5\rangle_{S,A}$ states, are used in truly chiral molecules. These states 
should have reasonably strong dipolar coupling with the $|3-4\rangle_{L,D}$ 
states, from the ground electronic surface, so that the light intensities 
are not too high for parasitic processes to take place.  Then the only 
requirement is for the process to be over before a $T_1$ type collisional 
relaxation of the states takes place. Recently, we have successfully applied
this approach on the permanently chiral molecule, that of 1,3-dimethylallene.

In experiments, we can tune the laser parameters to match the molecular 
parameters, until the conversion process becomes effective. We can, for 
example, continuously check the product by sensitively testing its circular 
birefringence \cite{circular}. The method can be applied even if the 
conversion is not perfect from various reasons, like when several 
molecular states are initially populated at higher temperatures. Then 
the purification can be completed by repeating the described process 
several times with the {\it same} set of pulses.  We just always let 
the system relax, so that the process starts with reasonable populations 
on the same initial states.

We believe that the new methodology presented here can largely influence 
the science and technology of chiral molecules purification, and lead to 
applications in organic chemistry, biochemistry and drug industry.

\vspace{1mm}
\noindent
We acknowledge many discussions with P. Brumer and E. Frishman. This project
was supported by the Minerva Foundation, GIF, the EU IHP program
HPRN-CT-1999-00129, the Office of Naval Research, USA, and the Swiss
Friends of the Weizmann Institute.



\begin{references}
\bibitem{STATES}
The $|1\rangle$, $|3\rangle$, $|4\rangle$ and $|5\rangle$ states correspond to
  torsional states with 0, 2, 3, 5 vibrational quanta, respectively. The
  $|2\rangle$ state is the first excited state of the asymmetric $S-D$
  stretching mode.
\end{references}
\end{document}